\documentclass{pasj00}

\SetRunningHead{A. Sekiguchi et al.}{Suzaku Observation of $\eta$ Car}

\Received{\today}
\Accepted{//}

\begin{document}
\title{Super-hard X-Ray Emission from $\eta$ Carinae Observed with Suzaku}
\author{
Akiko~\textsc{Sekiguchi},\altaffilmark{1,2}
Masahiro~\textsc{Tsujimoto}\altaffilmark{1,3}\thanks{Chandra Fellow}
Shunji~\textsc{Kitamoto}\altaffilmark{4},
Manabu~\textsc{Ishida}\altaffilmark{1}, 
Kenji~\textsc{Hamaguchi}\altaffilmark{5,6},\\
Hideyuki~\textsc{Mori}\altaffilmark{1}, and
Yohko~\textsc{Tsuboi}\altaffilmark{7}
}
\altaffiltext{1}{Japan Aerospace Exploration Agency, Institute of Space and
Astronautical Science,\\3-1-1 Yoshino-dai, Sagamihara, Kanagawa 229-8510}
\altaffiltext{2}{School of Physical Science, Space and Aeronautical Science, The
Graduate University for Advanced Studies\\Hayama-cho, Miura-gun, Kanagawa 240-0193}
\email{sekiguti@astro.isas.jaxa.jp}
\altaffiltext{3}{Department of Astronomy \& Astrophysics, Pennsylvania State University,\\
525 Davey Laboratory, University Park, PA 16802, USA}
\altaffiltext{4}{Department of Physics, Rikkyo University, 3-34-1 Nishi-Ikebukuro,
Toshima-ku, Tokyo 171-8501}
\altaffiltext{5}{CRESST and X-ray Astrophysics Laboratory NASA Goddard Space Flight Center, Greenbelt, MD 20771, USA}
\altaffiltext{6}{Department of Physics, University of Maryland, Baltimore County, 1000
Hilltop Circle, Baltimore, MD 21250, USA}
\altaffiltext{7}{Department of Science and Engineering, Chuo University, 1-13-27 Kasuga, Bunkyo-ku, Tokyo 112-8551}
\maketitle

\begin{abstract}
 We present the Suzaku results of $\eta$ Carinae in the 5--50~keV range conducted twice
 around the apastron in 2005 August for 50~ks and in 2006 February for 20~ks. The X-ray
 Imaging Spectrometer (XIS) produced hard (5--12~keV) band spectra, resolving K shell
 lines from highly ionized Fe and Ni. The Hard X-ray Detector yielded a significant
 detection in the super-hard (15--50~keV) band, which is uncontaminated by near-by
 sources. We constrained the temperature of the optically-thin thermal plasma emission
 dominant in the hard band as 3--4~keV using the K-shell line features with the XIS. We
 found significant excess emission above the thermal emission in the super-hard band
 with the PIN, confirming the previous INTEGRAL ISGRI report. The entire 5--50~keV
 spectra were fitted by a combination of a thermal plasma model plus a flat power-law or
 a very hot thermal bremsstrahlung model for the excess emission. No significant change
 of the excess emission was found at different epochs within the systematic and
 statistical uncertainties and no flare-like flux amplification was seen in the hard
 band, indicating that the excess emission is a steady phenomenon. We argue that the
 super-hard emission is attributable to the inverse Compton of stellar UV photons by
 non-thermal electrons or to the thermal bremsstrahlung of very hot plasma, and not to
 the bremsstrahlung by non-thermal electrons colliding with cold ambient matter.
\end{abstract}

\section{Introduction}
\begin{table*}[htbp]
 \begin{center}
  \caption{Observation log.}\label{t1}
  \begin{tabular}{lcccccccc}
   \hline
   Number & Sequence & \multicolumn{2}{c}{Observation start} & \multicolumn{2}{c}{Observation end} & \multicolumn{2}{c}{Exposure (ks)} & Orbital \\
          & number   & Date & Time                           & Date & Time                         & XIS & HXD & phase$^{*}$ \\     
   \hline
   First & 100012010 & 2005-08-29 & 01:48 & 2005-08-30 & 01:30 & 50 & 53 & 1.39 \\
   Second  & 100045010 & 2006-02-03 & 09:59 & 2006-02-03 & 22:45 & 21 & 17 & 1.47 \\
   \hline
   \multicolumn{7}{@{}l@{}}{\hbox to 0pt{\parbox{170mm}{\footnotesize
   \par\noindent
   \footnotemark[$*$]
   The orbital phase 1.0 indicates the onset of the Cycle 11 minimum at the Julian date
   2452819.8 d \citep{damineli08}.
   }\hss}}
  \end{tabular}
 \end{center}
\end{table*}

Colliding wind binaries (CWBs), a binary system comprised of two early-type stars with
stellar winds at high mass loss rates and velocities, are the brightest class of stellar
hard X-ray emitters. Theoretical interpretations are given for their luminous hard
X-rays as the thermal emission from high-temperature plasma produced by strong shocks
due to the colliding winds \citep{luo90,stevens92,usov92}. The interpretation is
supported by the X-ray modulation observed along the orbital motion in some CWBs
\citep{stevens96,ishibashi99,zhekov00}.

CWBs are also expected to be an agent of cosmic particle acceleration, which is
evidenced by the presence of non-thermal radio emission from e.\,g., WR\,140
\citep{white95}. Charged particles with an energy distribution deviating from the
Maxwellian distribution give rise to the X-ray emission harder than the thermal emission
dominant below 10~keV, thus the presence and the process of particle acceleration in
CWBs can be best studied in the super-hard X-ray band. For convenience, we refer to the
5--12~keV range as ``hard'' X-ray band, while the $>$15~keV range as ``super-hard''
X-ray band.

The investigation of the super-hard emission from CWBs is important to unveil their
enigmatic nature, and possibly gives a clue to understand the origin of $\gamma$-ray
emission reported from some massive star clusters containing CWBs
\citep{aharonian07}. We anticipate a great leap in our comprehension of non-thermal
phenomena in CWBs with the advent of X-ray imaging telescopes in the super-hard band,
which will be deployed in the next generation X-ray satellites such as Astro-H
\citep{takahashi08}. At present, however, our tool is limited to non-imaging or
coded-mask techniques in this band, leaving $\eta$ Carina the only practical target for
this subject.

$\eta$ Car is a luminous blue variable in the Carina nebula at a distance of 2.3~kpc
\citep{davidson97}. It is considered to constitute a binary system with another massive
star with an orbital period of 2022.7$\pm$1.3 d \citep{damineli08} in a highly eccentric
orbit \citep{nielson07}. In this paper, we use the orbital phase such that the start of
the Cycle 11 minimum is phase 1.0 on the Julian date of 2452819.8 d \citep{damineli08}
or 2003 June 29.3. $\eta$ Car is one of the brightest stellar hard X-ray sources, and
the hard X-ray modulation along the proposed ephemeris endorses the colliding-wind
origin of the emission \citep{ishibashi99}.

The development of the X-ray light curve through an orbital phase is as follows; it
experiences an eclipse from phase 0 to 0.03 (we call ``minimum''), recovers its
brightness spending $\sim$0.02 phases (``recovery''), and gradually increases the flux
for the rest of the phase (``steady increase'') to reach the brightest point
(``maximum''), which is terminated by the repeated eclipse in the next phase
\citep{ishibashi99,pittard02}. The periastron and the apastron are supposed to occur
around phase 0.0 and 0.5, respectively (e.\,g., \cite{fernandez-lajus09}). During the
development, $\eta$ Car shows episodic flares \citep{ishibashi99,hamaguchi07a}. The
spectrum of $\eta$ Car below 10~keV has been described by complex models comprised of
several thin-thermal plasma components with different temperatures, but a single plasma
component of a 3--4~keV temperature dominates the hard band emission
\citep{tsuboi97,corcoran98}.

\medskip

X-ray studies of $\eta$ Car in the super-hard band were conducted by two observatories
to date. One is by BeppoSAX using a combination of a gas scintillation proportional
counter (MECS; Medium Energy Concentrator Spectrometer) below 10~keV and a non-imaging
phoswich scintillators (PDS; Phoswich Detection System) above 15~keV. \citet{viotti02}
and \citet{viotti04} constrained the parameters of the 3--4~keV plasma component using
the hard band data by the MECS, and extrapolated the model to claim that the PDS data
contain excess emission upon the 3--4~keV thermal emission. The other is by the
INTErnational Gamma-Ray Astrophysics Laboratory (INTEGRAL) using a coded-mask imager
(ISGRI; the INTEGRAL Soft Gamma-Ray Imager) in the 15~keV--1~MeV range. \citet{leyder08}
unambiguously confirmed the super-hard emission claimed by the BeppoSAX studies, showing
that the ISGRI image has a significant signal at the position of $\eta$ Car.

When it comes to characterizing the spectrum of the super-hard excess emission, the
results produced by the two observatories have a limited use with some concerns. The PDS
is a non-imaging instrument with a wide field of view of \timeform{1.3$^\circ$} in the
full width at the half maximum (FWHM), thus the contamination by other sources is
inevitable. Indeed, as was pointed out by \citet{leyder08}, two neighboring super-hard
sources (IGR\,J10447--6027 and 1E\,1048.1--5937) are within the PDS field of view, which
significantly contribute to the PDS spectrum of $\eta$ Car. The ISGRI resolved super
hard emission from $\eta$ Car and is free from the contamination by these two
sources. However, its spectrum does not have rich statistics with only three spectral
bins for a total of a 1~Ms exposure. The flux variation cannot be tested. Also, the
excess emission was examined against the extrapolated model of the 3--4~keV thermal
emission obtained by the MECS at a different epoch, which may not be very appropriate
given the variable nature of the hard X-rays of $\eta$ Car.

The detection and the characterization of the excess emission in the super-hard band
heavily depend on the accurate characterization and subtraction of the 3--4~keV thermal
emission. The examination of the excess emission is straight-forward, but it requires a
careful treatment since the extrapolation of the 3--4~keV thermal model needs to be
stretched for nearly an order of magnitude in the energy range.

With these concerns, it is not surprising that the BeppoSAX and the INTEGRAL data
resulted in a serious disagreement with each other for the super-hard excess
emission. While the former claims the power-law model of a photon index of
2.42$\pm$0.04, the latter claims 1$\pm$0.4. Moreover, the flux estimates are
inconsistent significantly by a factor of 3. It is unclear, in the first place, whether
the super-hard excess emission is of a thermal or a non-thermal origin.

\medskip

To mitigate some of these concerns in the previous studies, we present the 5--50~keV
results of Suzaku observations conducted in 2005 August and 2006 February. The
non-imaging detector in the super-hard band on board Suzaku has a narrower field of view
than that of the PDS and is free from the contamination by the two neighboring super
hard sources. The X-ray charge coupled devices (CCDs) sensitive below 12~keV have good
and well-calibrated responses in the hard band. Their spectroscopic capability
outperforms that of the MECS, resolving emission lines in the Fe and Ni K complex in the
6.0--8.5~keV band and thus enabling an accurate temperature determination of the
3--4~keV plasma sufficient to extrapolate toward the super-hard band.

We present the spectral models to explain the 5--50~keV emission of $\eta$ Car at two
epochs around the apastron. We examine the thermal or non-thermal nature of the super
hard excess emission and investigate its flux variability. The results will set a firm
basis to study the development of the super-hard emission in an entire orbital cycle,
which we plan to conduct in a separate paper with the new data obtained at the end of
2008 during the latest X-ray maximum and minimum.

\section{Observations and Data Reduction}
\begin{figure*}[hbtp]
 \begin{center}
  \FigureFile(180mm,85mm){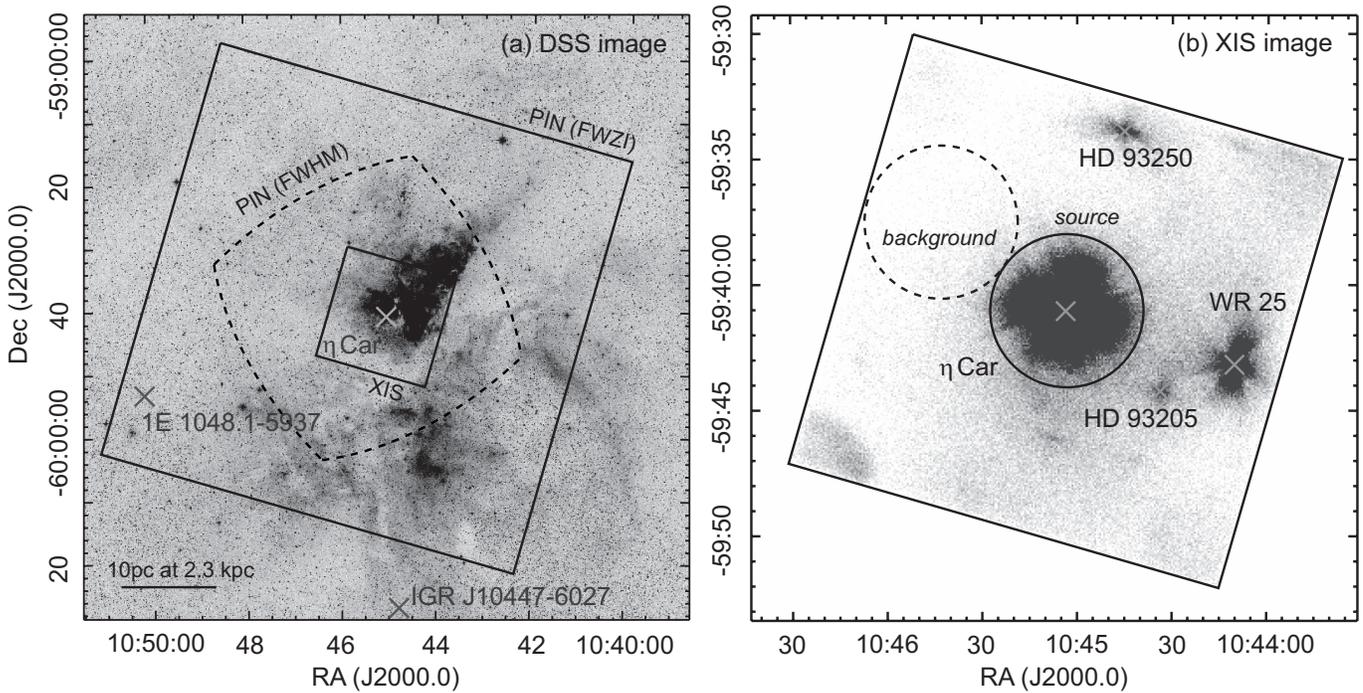}
 \end{center}
 \caption{(a) Fields of view of the XIS (small solid square) and the PIN (large solid
 square for FWZI and dashed region for FWHM) in the first observation overlaid on an
 optical image by the Digitized Sky Survey around $\eta$ Car. The positions of the
 INTEGRAL super-hard sources are shown \citep{leyder08} with crosses. (b) XIS image in
 the first observation. The events obtained by the four CCDs are merged. The positions
 and the names of $\eta$ Car and the neighboring resolved bright X-ray sources are
 labeled. The solid and the dashed circles represent the source and the background
 extraction regions, respectively. The X-ray events in the corners stem from the
 calibration sources.}\label{f1}
\end{figure*}

The Suzaku satellite \citep{mitsuda07} observed the Carina nebula multiple times since
its launch. We present the results of $\eta$ Car in the first two observations on 2005
August 29 and 2006 February 3 in this paper, during which the object was around the
apastron at the orbital phases of 1.39 and 1.47, respectively (table~\ref{t1}). The two
observations were centered at (RA, Dec) $=$ (\timeform{10h45m03.6s},
\timeform{-59D41'04''}) in the equinox J2000.0 with nearly the opposite position
angles. Using the first data set, \citet{hamaguchi07b} studied the soft diffuse emission
around $\eta$ Car.

Suzaku has a low earth orbit with an altitude of $\sim$570~km. The observatory has two
working instruments; the X-ray Imaging Spectrometer (XIS; \cite{koyama07}) and the Hard
X-ray Detector (HXD; \cite{takahashi07,kokubun07}), which respectively covers the energy
range of 0.2--12~keV and 10--600~keV.

The XIS is an imaging spectrometer using X-ray CCDs with a format of 1024$\times$1024
pixels. Four CCD cameras are installed in the focal planes of four X-ray telescope
modules \citep{serlemitsos07} coaligned with each other. Three of the four cameras (XIS
0, 2, and 3) are front-illuminated (FI) devices and the remaining one (XIS1) is
back-illuminated (BI). The FI and the BI CCDs have superior sensitivity to each other in
the hard and the soft bands, respectively. The energy resolution, which is subject to
degradation in the orbit, is $\sim$140~eV and $\sim$160~eV at 5.9~keV in FWHM
respectively for the first and the second observations. The absolute energy gain is
accurate to $\sim$0.2\% at 5.9~keV. Together with the XRT, the XIS covers a field of
view of 18\arcmin$\times$18\arcmin\ with a pixel scale of 1\arcsec~pixel$^{-1}$. The
half-power diameter of the XRT is 1\farcm8--2\farcm3, which is almost independent of
energy.

The HXD is a non-imaging spectrometer, consisting of several components responsible for
different energy ranges. In this paper, we use the PIN component sensitive at 10--70~keV
with an FWHM energy resolution of 3.0~keV. The PIN detector is composed of 64 Si PIN
diodes installed at the bottom of 8$\times$8 well-type collimators surrounded by GSO
scintillators. The fine collimators within each well restrict the field of view. The
effective area monotonically decreases as an increasing distance from the field center,
which is characterized by the FWHM view of $\sim$34\arcmin\ or the full width at zero
intensity (FWZI) view of 70\arcmin\ square (figure~\ref{f1}). The center of the PIN
field is 3\farcm5 offset from that of the the XIS field. Thanks to the surrounding
anti-coincidence scintillators, the narrow field of view, and the low background
environment in the orbit, the PIN achieves unprecedented sensitivity in the super-hard
band.

In both observations, the XIS is operated in the normal clocking mode with a frame time
of 8~s. We used the cleaned event lists produced by the pipeline processing version
2.0. The net exposure times are compiled in table~\ref{t1}. We used the HEASoft version
6.5.1\footnote{See http://heasarc.gsfc.nasa.gov/docs/software/lheasoft/ for detail.} for
the data reduction and Xspec version 11.3.2\footnote{See
http://heasarc.gsfc.nasa.gov/docs/xanadu/xspec/index.html for detail.} for the X-ray
spectral analysis.

\section{Analysis}
\subsection{Constructing the Spectra}
\subsubsection{XIS}
Figure~\ref{f1} (b) shows the XIS image in the first observation. Three point sources
are recognized besides $\eta$ Car: HD\,93250, HD\,93205, and WR\,25, all of which are
massive stars in this region. The underlying diffuse emission, which is a combination of
unresolved point sources and the soft extended emission
\citep{evans03,evans04,colombo03,hamaguchi07b}, is noticeable particularly in the
western half of the image. The second observation produced a similar image.

In order to construct the XIS spectra, we extracted the source and the background events
from the solid and the dashed circles in figure~\ref{f1} (b), respectively. The source
extraction region is a 3\arcmin\ radius circle containing $\sim$95\% of photons from a
point-like source at the circle center. The background region is selected from a region
devoid of bright emission. The pile-up is negligible. The wings of other point sources
and the underlying diffuse emission contribute to the background, which is negligible in
our targeted energy range. This is illustrated by the fact that the emission above 5~keV
at the X-ray minimum of $\eta$ Car, which gives the upper limit to the contamination by
irrelevant emission, is roughly two orders of magnitude smaller than that in the X-ray
bright phase \citep{hamaguchi07a}.

\begin{figure*}
 \begin{center}
  \FigureFile(180mm,85mm){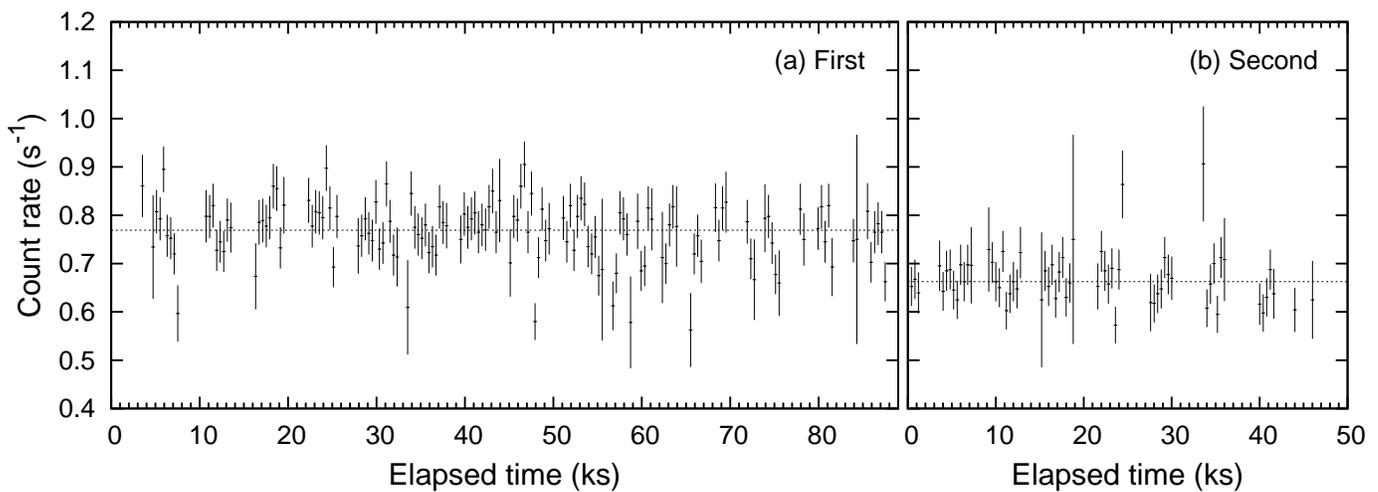}
 \end{center}
 \caption{XIS 5--10~keV light curve binned with 400~s~bin$^{-1}$ in (a) the first and
 (b) the second observations. The best-fit constant model is shown with the dashed
 lines. The origin of the abscissa is the start of the observation.}\label{f8}
\end{figure*}

We constructed hard band light curves of the count rate during the 50 and 21~ks
exposures in the first and the second observations, respectively, and found no
significant flux changes such as flare-like variability (figure~\ref{f8}). For the
spectral modeling, we merged the spectra obtained by the three FI devices with nearly
identical responses. The merged FI spectrum has a good signal up to $\sim$12~keV, while
the BI spectrum is effective up to $\sim$9~keV. We generated the redistribution matrix
functions and the ancillary response files using the \texttt{xisrmfgen} and
\texttt{xissimarfgen} tools \citep{ishisaki07}, respectively.

\subsubsection{PIN}
Figure~\ref{f1} (a) shows the field of view of the PIN detector. One of the two INTEGRAL
sources (IGR\,J10447--6027) is outside of the FWZI view.  The other (1E\,1048.1--5937)
is close to the edge of the view in both observations, at which the effective area
decreases to less than 5\% of the value at the field center. \citet{leyder08} claimed
the INTEGRAL ISGRI count rate of 0.16~s$^{-1}$ for $\eta$ Car and 0.09~s$^{-1}$ for
1E\,1048.1--5937. Assuming that 1E\,1048.1--5937 does not vary drastically among the
INTEGRAL and the two Suzaku observations, the contamination by this source is negligible
in the PIN spectrum.

We constructed the background spectrum, which is composed mostly of the non X-ray
background (NXB) with some contribution by the Cosmic X-ray Background (CXB). The
Galactic Ridge X-ray emission is negligible toward $\eta$ Car at a galactic longitude of
$\sim$290$^{\circ}$. For the NXB, we used the simulated background events distributed by
the instrument team, which reproduces the NXB spectrum at an accuracy of
1.3\%\footnote{See
http://www.astro.isas.jaxa.jp/suzaku/doc/suzakumemo/\\suzakumemo-2008-03.pdf and
http://heasarc.gsfc.nasa.gov/\\docs/suzaku/analysis/pinbgd.html for detail.}. For the
CXB, we assumed the spectral model derived from HEAO-1 observations \citep{boldt87}. We
produced a CXB spectrum by convolving the detector angular and energy responses with the
simulated emission of the assumed model and the spatially uniform distribution. The
resultant NXB and CXB data were merged to construct the PIN background spectrum.

For the spectral modeling, we used the latest detector response for a point source at
the XIS field center (\texttt{ae\_hxd\_pinxinome1\_20080129.rsp}). At this position, the
relative flux among the XIS devices and the PIN is calibrated at an accuracy of
$\sim$1\%\footnote{See
http://www.astro.isas.jaxa.jp/suzaku/doc/suzakumemo/\\suzakumemo-2007-11.pdf for
detail.}.

\subsection{Modeling the Spectra}
\subsubsection{Fitting below 10~keV}\label{s321}
\begin{figure}
 \begin{center}
  \FigureFile(85mm,85mm){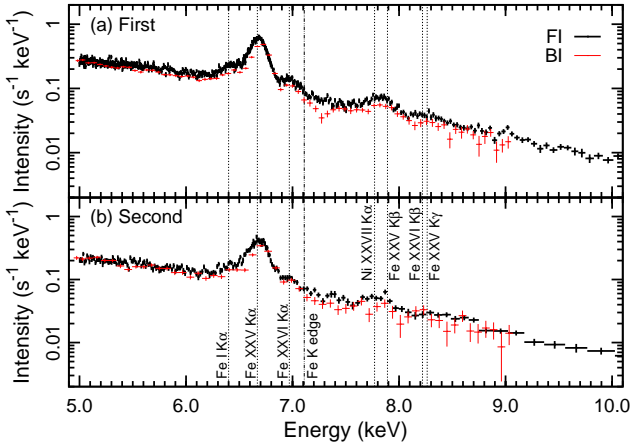}
 \end{center}
 \caption{XIS spectra in the 5--10~keV range in (a) the first and (b) the second
 observations. The FI and BI spectra are shown with black and red error bars,
 respectively, with bold face symbols for the FI spectra. The conspicuous emission lines
 as well as the Fe K edge are labeled.}\label{f2}
\end{figure}

With the background-subtracted spectra and the response files both for the XIS and the
PIN, we are now ready for the spectral modeling. It is not a wise decision to start
fitting the two spectra simultaneously; the statistics of the XIS spectrum overwhelms
that of the PIN spectrum, making the PIN data ignored when fitted at the equal
weight. We therefore took a more step-wise approach described below.

Figure~\ref{f2} shows the XIS spectra in the 5--10~keV range, in which we can resolve
emission lines from highly ionized Fe and Ni as well as hard continuum emission
extending beyond 10~keV. These are the characteristics of the emission arising from a
thin-thermal plasma of several keV temperatures. In addition to the thermal features, we
can also find an emission line at $\sim$6.4~keV due to the K$\alpha$ emission of
quasi-neutral Fe (hereafter called Fe\emissiontype{I}) and an absorption feature at
7--8~keV due to the K-shell photoelectric absorption edge by Fe\emissiontype{I}.

The rich set of thermal features in our XIS spectra enables us to constrain the plasma
temperature stringently, in particular, by using the intensity ratio between the
Fe\emissiontype{XXV} K$\alpha$ line at 6.7~keV and Fe\emissiontype{XXV} K$\beta$ line at
7.9~keV and that between Fe\emissiontype{XXV} K$\alpha$ line at 6.7~keV and
Fe\emissiontype{XXVI} K$\alpha$ line at 7.0~keV.

\begin{table}
 \begin{center}
  \caption{Best-fit model for 6.0--8.5~keV spectra.}\label{t2}
 \end{center}
\begin{tabular}{lcc}
 \hline
 \hline 
 Parameters$^*$ & First$^\dagger$ & Second$^\dagger$ \\
 \hline
 $k_{\rm{B}}T^{\rm{(th)}}$ (keV)   & 3.9$^{+0.1}_{-0.1}$    & 4.1$^{+0.3}_{-0.3}$\\
 $Z_{\rm{Fe}}^{\rm{(th)}}$ (solar) & 0.47$^{+0.01}_{-0.02}$ & 0.45$^{+0.02}_{-0.03}$ \\
 $Z_{\rm{Ni}}^{\rm{(th)}}$ (solar) & 1.4$^{+0.2}_{-0.2}$    & 0.9$^{+0.4}_{-0.4}$ \\
 $F_{\rm{X}}^{\rm{(th)}}$ ($10^{-11}$~erg~s$^{-1}$~cm$^{-2}$) & $2.90^{+0.02}_{-0.03}$ & $2.23^{+0.03}_{-0.12}$ \\
 \hline
 $E^{\rm{(gau)}}$ (keV)  & 6.42$^{+0.01}_{-0.01}$       & 6.43$^{+0.01}_{-0.01}$ \\
 $I^{\rm{(gau)}}$ ($10^{-5}$~s$^{-1}$~cm$^{-2}$) & $8.5^{+0.6}_{-0.6}$  & $6.1^{+0.9}_{-0.9}$  \\
 \hline
 $E^{\rm{(edge)}}$ (keV) & 7.11$^{+0.07}_{-0.06}$ & 6.97$^{+0.15}_{-0.10}$ \\
 $\tau^{\rm{(edge)}}$    & 0.12$^{+0.03}_{-0.03}$ & 0.12$^{+0.08}_{-0.04}$ \\
 \hline
 $\chi^2$/d.o.f. & 346.2/323 & 143.9/138 \\
 \hline
 \multicolumn{3}{@{}l@{}}{\hbox to 0pt{\parbox{85mm}{\footnotesize
 \par\noindent
 \footnotemark[$*$] The superscripts of the parameters represent the spectral component
 that they belong to: ``(th)'' for the thin-thermal plasma, ``(gau)'' for the Gaussian
 line, and ``(edge)'' for the photoelectric absorption
 model. $F_{\rm{X}}^{\rm{(th)}}$ is derived in 5--10~keV.
 \par\noindent
 \footnotemark[$\dagger$] The best-fit values for the first and the second
 observations. The ranges indicate the 90\% statistical uncertainty.
 }\hss}}
\end{tabular}
\end{table}

\begin{figure*}[hbt]
 \begin{center}
  \FigureFile(180mm,85mm){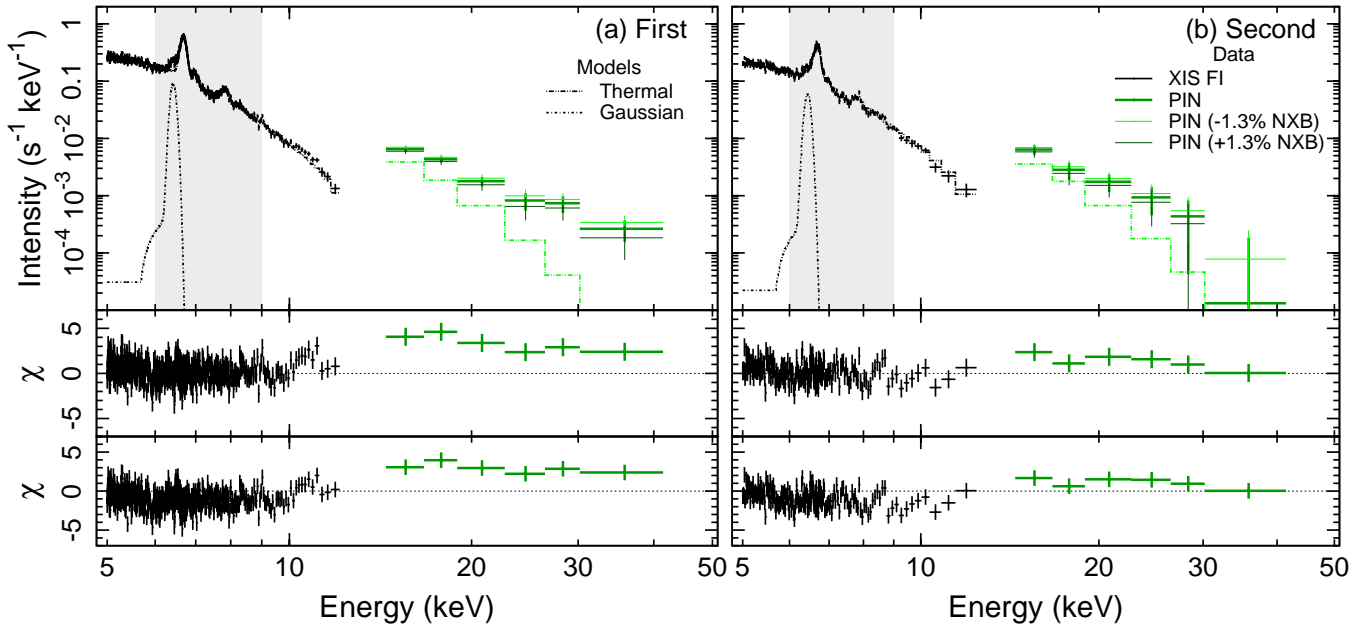}
 \end{center}
 \caption{5--50~keV spectrum compared to the best-fit model for the 6.0--8.5~keV (the
 gray shaded range) spectrum in (a) the first and (b) the second observations. The top
 panels show the data and each spectral component of the model (see the legend in the
 top left panel). For the XIS spectra, we plot only the FI spectrum. For the PIN
 spectra, three spectra are shown with different darkness of green colors. Each
 represents the data constructed with three NXB models; i. e., the distributed model and
 its scaled models by the systematic uncertainty of $\pm$1.3\% (see the legend at the
 top right panel). The middle panels show the residuals of the data from the best-fit
 6.0--8.5~keV model. The bottom panels show the residuals of the data from the model
 using the maximum allowed values of $k_{\rm{B}}T^{\rm{(th)}}$ and
 $F_{\rm{X}}^{\rm{(th)}}$ of the best-fit 6.0--8.5~keV model
 (table~\ref{t2}).}\label{f3}
\end{figure*}

To fully exploit the XIS capability, we fitted the 6.0--8.5~keV range spectra, in which
we can find most of the emission lines responsible for the plasma temperature
determination. We applied the model composed of a thin-thermal plasma component
(\texttt{vmekal}; \cite{mewe85,mewe86,kaastra92,liedahl95}) for the thermal features and
a Gaussian line component for the Fe\emissiontype{I} K$\alpha$ line. We attenuated the
model by an interstellar extinction model (\texttt{wabs}; \cite{morrison83}) as well as
an additional photoelectric absorption edge to account for the Fe K edge feature. The
free parameters in the fitting are the plasma temperature ($k_{\rm{B}}T^{\rm{(th)}}$),
the Fe and Ni abundance ($Z_{\rm{Fe}}^{\rm{(th)}}$ and $Z_{\rm{Ni}}^{\rm{(th)}}$,
respectively), and the flux ($F_{\rm{X}}^{\rm{(th)}}$) in the 5--10~keV band for the
thin-thermal plasma component, the line energy ($E^{\rm{(gau)}}$) and the intensity
($I^{\rm{(gau)}}$) for the Gaussian line component, and the edge energy
($E^{\rm{(edge)}}$) and the optical depth ($\tau^{\rm{(edge)}}$) for the photoelectric
absorption edge component. Here, the superscripts of the parameters represent the
spectral component that they belong to. The amount of extinction was fixed to
1.7$\times$10$^{23}$~cm$^{-2}$, which is a typical value derived in the hard band for
the colliding-wind plasma of $\eta$ Car \citep{hamaguchi07a}. The intrinsic width of the
Gaussian line was fixed to 0~eV. The linearity coefficient between the pulse height and
the energy was adjusted to compensate for a possible systematic uncertainty.

We obtained statistically-acceptable fits both for the first and the second observations
(table~\ref{t2}). The plasma temperature ($k_{\rm{B}}T^{\rm{(th)}}$) and the flux
($F_{\rm{X}}^{\rm{(th)}}$) were determined at an accuracy of $\sim$5\%. The flux between
the two observations changed significantly by $\sim$20\%. It is interesting to note
that, while $Z_{\rm{Fe}}^{\rm{(th)}}$ and $Z_{\rm{Ni}}^{\rm{(th)}}$ are consistent
between the two observations, $Z_{\rm{Ni}}^{\rm{(th)}}$ is larger than
$Z_{\rm{Fe}}^{\rm{(th)}}$ by a factor of a few; its cause is not clear.

\subsubsection{Extending beyond 10~keV}\label{s322}
We extrapolated the best-fit 6.0--8.5~keV model beyond 10~keV and compared to the PIN
spectrum in order to examine the excess emission in the super-hard band. The top panels
in figure~\ref{f3} show the XIS and the PIN spectra below 12~keV and above 15~keV,
respectively, in comparison with the extrapolated best-fit model in the 6.0--8.5~keV
range. The middle panels show the residuals, in which we can see a consistent positive
residual for the PIN spectrum in both observations. This illustrates the presence of the
excess emission beyond 15~keV.

We tested the claim for two major sources of systematic uncertainty. One is the
systematic uncertainty in the NXB model of the PIN data, which constitutes the majority
of the background signal. We constructed the background-subtracted PIN spectra for the
scaled NXB models by $\pm$1.3\%, which is the accuracy of the model, and plotted them
against the extrapolated model in different darkness of green colors in the top panels
of figure~\ref{f3}. For the increased NXB model, we still find consistent positive
residuals beyond 10~keV. The other is the statistical uncertainty in the best-fit
6.0--8.5~keV model. In order to examine the upper bound of the extrapolated model, we
replaced the best-fit $k_{\rm{B}}T^{\rm{(th)}}$ and $F_{\rm{X}}^{\rm{(th)}}$ values with
their maximum allowed values within 90\% statistical uncertainty. The modified model was
extrapolated and compared to the data. The residuals are shown in the bottom panels in
figure~\ref{f3}, which again show consistent positive residuals of the data above the
extrapolated model.

\begin{figure*}[hbt]
 \begin{center}
  \FigureFile(180mm,85mm){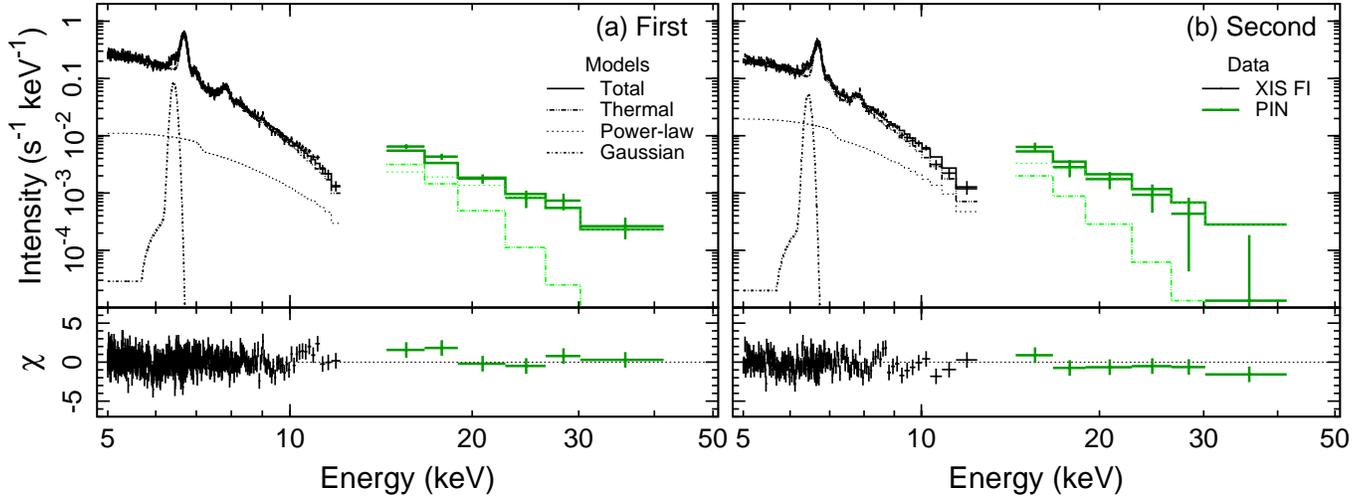}
 \end{center}
 \caption{5--50~keV spectrum and the best-fit model. A power-law model is added to
 explain the super-hard excess emission. The top panels show the data and the best-fit
 spectra and their components (see the legend in the top left panel). XIS FI and PIN
 data are shown with black and green colors, respectively. The best-fit parameters of
 the models are compiled in table~\ref{t3}.}\label{f4}
\end{figure*}

\begin{figure*}[hbt]
 \begin{center}
  \FigureFile(180mm,85mm){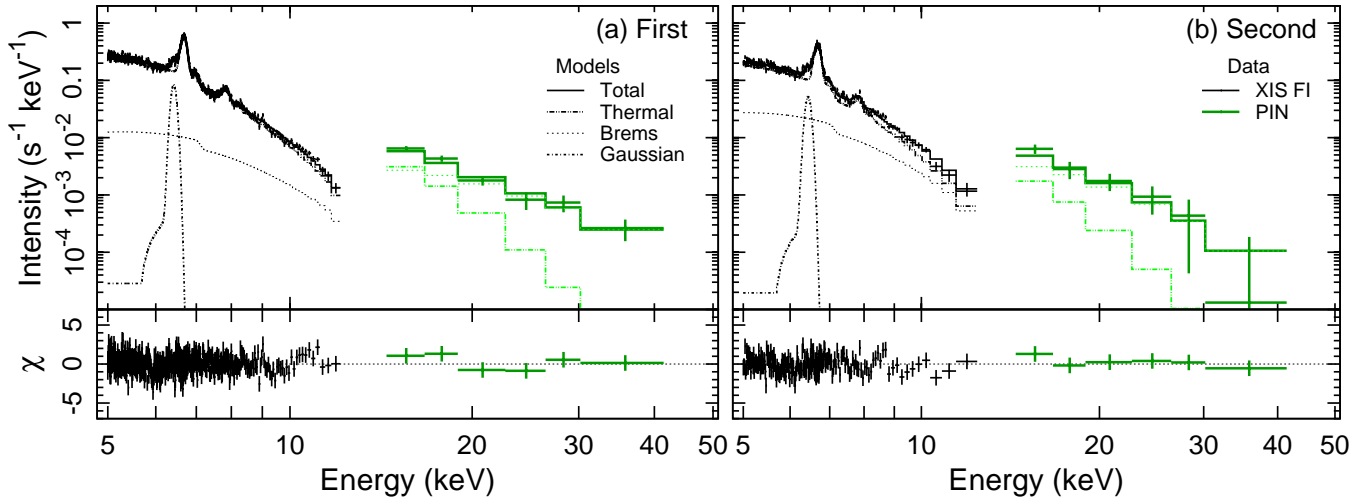}
 \end{center}
 \caption{5--50~keV spectrum and the best-fit model. A thermal bremsstrahlung model is
 added to explain the super-hard excess emission. The symbols and colors follow
 figure~\ref{f4}. The best-fit parameters of the models are compiled in
 table~\ref{t4}.}\label{f5}
\end{figure*}

\subsubsection{Fitting the Entire Band}\label{s323}
We established the presence of the excess emission in the super-hard band and now
proceed to the spectral modeling in the entire 5--50~keV range using both the XIS and
the PIN data. For the hard emission, we used the same model for the 6.0--8.5~keV fitting
procedure in subsection~\ref{s321} with a slight modification that the
$Z_{\rm{Fe}}^{\rm{(th)}}$ and $Z_{\rm{Ni}}^{\rm{(th)}}$ are tied between the two
observations. Upon this thermal emission, we added another component to account for the
super-hard excess emission. Two models were employed; a power-law model and a thermal
bremsstrahlung. The free parameters in the additional components are the power-law
photon index ($\Gamma^{\rm{(pl)}}$) and the flux $F_{\rm{X}}^{\rm{(pl)}}$ in the
10--50~keV range for the power-law model and the electron temperature
($k_{\rm{B}}T^{\rm{(br)}}$) and the 10--50~keV flux ($F_{\rm{X}}^{\rm{(br)}}$) for the
bremsstrahlung model. Both yielded an acceptable fit. The best-fit values are tabulated
in tables~\ref{t3} and \ref{t4}, while the data and the best-fit model are shown in
figures~\ref{f4} and \ref{f5}, respectively, for the additional power-law and
bremsstrahlung models.

We assessed the systematic uncertainty in the derived parameters for the two major
sources of such uncertainty. First, for the PIN NXB model uncertainty, we repeated the
5--50~keV fitting procedure using the scaled NXB models by $\pm$1.3\%. For the power-law
model, we found that this brings systematic uncertainties of ($\Gamma$,
$F_{\rm{X}}^{\rm{(pl)}}$/10$^{-11}$~erg~s$^{-1}$~cm$^{-2}$) $\sim$ (0.2, 0.3) for the
first and (0.1, 0.3) for the second observations. For the bremsstrahlung model, the
systematic uncertainty for $F_{\rm{X}}^{\rm{(bb)}}$/$10^{-11}$~erg~s$^{-1}$~cm$^{-2}$ is
$\sim$0.6 and $\sim$ 0.5 in the first and the second observations, respectively. The
resultant systematic uncertainty of $k_{\rm{B}}T^{\rm{(bb)}}$ is so large that it makes
$k_{\rm{B}}T^{\rm{(bb)}}$ practically unconstrained in the first observation and
$>$16~keV in the second observation. Next, for the statistical uncertainty in the
best-fit 3--4~keV thermal model parameters, we replaced the best-fit set of
$k_{\rm{B}}T^{\rm{(th)}}$ and $F_{\rm{X}}^{\rm{(th)}}$ values with those of their
maximum and minimum allowed values within 90\% statistical uncertainty. This brings
systematic uncertainties of ($\Gamma$,
$F_{\rm{X}}^{\rm{(pl)}}$/10$^{-11}$~erg~s$^{-1}$~cm$^{-2}$) $\sim$ (0.02, 0.2) and
(0.01, 0.2) respectively for the first and the second observations in the power-law
model. For the bremsstrahlung model, the $k_{\rm{B}}T^{\rm{(bb)}}$~keV is unconstrained
with minima of $>$22~keV and $>$10~keV for the first and the second observations. The
resultant systematic uncertainty for
$F_{\rm{X}}^{\rm{(bb)}}$/$10^{-11}$~erg~s$^{-1}$~cm$^{-2}$ is 0.3 and 0.2, respectively.

\begin{table}
  \caption{Best-fit model for the 5--50~keV spectra using a power-law
  model for the super-hard excess emission.}\label{t3}
\begin{tabular}{lcc}
 \hline
 \hline 
 Parameters$^*$ & First$^\dagger$ & Second$^\dagger$ \\
 \hline
 $\Gamma^{\rm{(pl)}}$ & 1.30$^{+0.07}_{-0.14}$ & 1.51$^{+0.04}_{-0.15}$ \\
 $F_{\rm{X}}^{\rm{(pl)}}$ ($10^{-11}$~erg~s$^{-1}$~cm$^{-2}$) & 1.4$^{+0.2}_{-0.1}$ &  1.9$^{+0.2}_{-0.2}$ \\
 \hline
 $k_{\rm{B}}T^{\rm{(th)}}$ (keV)   & 3.61$^{+0.03}_{-0.03}$   & 3.29$^{+0.03}_{-0.03}$\\
 $Z_{\rm{Fe}}^{\rm{(th)}}$ (solar) & \multicolumn{2}{c}{0.47$^{+0.01}_{-0.01}$}    \\
 $Z_{\rm{Ni}}^{\rm{(th)}}$ (solar) & \multicolumn{2}{c}{1.5$^{+0.2}_{-0.2}$}     \\
 $F_{\rm{X}}^{\rm{(th)}}$ ($10^{-11}$~erg~s$^{-1}$~cm$^{-2}$) & 2.76$^{+0.02}_{-0.01}$ & 2.07$^{+0.02}_{-0.02}$ \\
 \hline
 $E^{\rm{(gau)}}$ (keV)  & 6.43$^{+0.01}_{-0.01}$       & 6.44$^{+0.01}_{-0.02}$ \\
 $I^{\rm{(gau)}}$ ($10^{-5}$~s$^{-1}$~cm$^{-2}$)  & 7.9$^{+0.5}_{-0.5}$  & 5.5$^{+0.8}_{-0.7}$  \\
 \hline
 $E^{\rm{(edge)}}$ (keV) & 7.12$^{+0.04}_{-0.04}$ & 7.14$^{+0.17}_{-0.10}$ \\
 $\tau^{\rm{(edge)}}$    & 0.16$^{+0.02}_{-0.02}$ & 0.14$^{+0.04}_{-0.04}$ \\
 \hline
 $\chi^2$/d.o.f. & 717.0/590 & 286.1/226 \\
 \hline
 \multicolumn{3}{@{}l@{}}{\hbox to 0pt{\parbox{85mm}{\footnotesize
 \footnotemark[$*$] The superscripts of the parameters represent the spectral component
 that they belong to: ``(pl)'' for the power-law, ``(th)'' for the thin-thermal plasma,
 ``(gau)'' for the Gaussian line, and ``(edge)'' for the photoelectric absorption
 model. $F_{\rm{X}}^{\rm{(th)}}$ is derived in 5--10~keV, while
 $F_{\rm{X}}^{\rm{(pl)}}$ is in 10--50~keV.
 \par\noindent
 \footnotemark[$\dagger$] The best-fit values for the first and the second
 observations. The ranges indicate the 90\% statistical uncertainty.
 }\hss}}
\end{tabular}
\end{table}

\begin{table}
 \caption{Best-fit model for the 5--50~keV spectra using a thermal bremsstrahlung model
 for the super-hard excess emission.}\label{t4}
\begin{tabular}{lcc}
 \hline
 \hline 
 Parameters$^*$ & First$^\dagger$ & Second$^\dagger$ \\
 \hline
 $k_{\rm{B}}T^{\rm{(br)}}$ (keV) & $>61$ & 18$^{+20}_{-8}$ \\
 $F_{\rm{X}}^{\rm{(br)}}$ ($10^{-11}$~erg~s$^{-1}$~cm$^{-2}$) & 1.7$^{+0.5}_{-0.4}$ &  1.2$^{+0.9}_{-0.8}$ \\
 \hline
 $k_{\rm{B}}T^{\rm{(th)}}$ (keV)   & 3.58$^{+0.07}_{-0.11}$ & 3.14$^{+0.28}_{-0.14}$\\
 $Z_{\rm{Fe}}^{\rm{(th)}}$ (solar) & \multicolumn{2}{c}{0.47$^{+0.01}_{-0.01}$} \\
 $Z_{\rm{Ni}}^{\rm{(th)}}$ (solar) & \multicolumn{2}{c}{1.5$^{+0.2}_{-0.2}$} \\
 $F_{\rm{X}}^{\rm{(th)}}$ ($10^{-11}$~erg~s$^{-1}$~cm$^{-2}$) & 2.74$^{+0.10}_{-0.12}$ & 1.95$^{+0.11}_{-0.18}$ \\
 \hline
 $E^{\rm{(gau)}}$ (keV)  & 6.43$^{+0.01}_{-0.01}$       & 6.43$^{+0.01}_{-0.02}$ \\
 $I^{\rm{(gau)}}$ ($10^{-5}$~s$^{-1}$~cm$^{-2}$) & 7.8$^{+0.3}_{-0.5}$  & 5.3$^{+0.8}_{-0.8}$  \\
 \hline
 $E^{\rm{(edge)}}$ (keV) & 7.12$^{+0.06}_{-0.04}$ & 7.1$^{+0.2}_{-0.1}$ \\
 $\tau^{\rm{(edge)}}$    & 0.16$^{+0.04}_{-0.03}$ & 0.14$^{+0.08}_{-0.06}$ \\
 \hline
 $\chi^2$/d.o.f. & 714.1/590 & 276.2/226 \\
 \hline
 \multicolumn{3}{@{}l@{}}{\hbox to 0pt{\parbox{85mm}{\footnotesize
 \footnotemark[$*$] The superscripts of the parameters represent the spectral component
 that they belong to: ``(br)'' for the bremsstrahlung, ``(th)'' for the thin-thermal plasma,
 ``(gau)'' for the Gaussian line, and ``(edge)'' for the photoelectric absorption
 model. $F_{\rm{X}}^{\rm{(th)}}$ is derived in 5--10~keV, while
 $F_{\rm{X}}^{\rm{(br)}}$ is in 10--50~keV.
 \par\noindent
 \footnotemark[$\dagger$] The best-fit values for the first and the second
 observations. The ranges indicate the 90\% statistical uncertainty or the
 lower limit. $k_{\rm{B}}T^{\rm{(br)}}$ is unconstrained in the first observation.
 }\hss}}
\end{tabular}
\end{table}

\section{Discussion}
\subsection{Comparison with the Previous Results}
\begin{figure}[hbt]
 \begin{center}
  \FigureFile(85mm,85mm){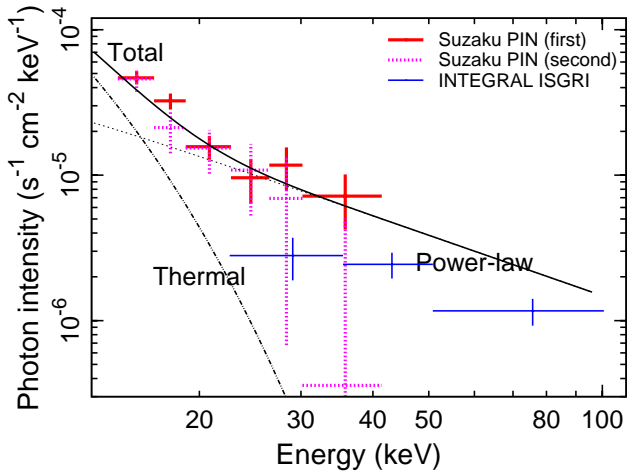}
 \end{center}
 \caption{super-hard spectra by the two Suzaku and the INTEGRAL observations. The
 best-fit model for the merged Suzaku spectrum is shown (dotted for the power-law,
 dashed-and-dotted for the thin-thermal, and solid for the total model).}\label{f6}
\end{figure}

\begin{table}
 \begin{center}
  \caption{Comparison with previous results.}\label{t5}
  \begin{tabular}{lccc}
   \hline
   \hline
   Observatory & $\Gamma^{*\dagger}$ & $F_{\rm{X}}^{*\dagger}$/10$^{-11}$ & Orbital\\
   &  & (erg~s$^{-1}$~cm$^{-2}$) & phase$^{\ddagger}$ \\
   \hline
   BeppoSAX      &  2.42$^{+0.04}_{-0.04}$ & 3.24 & 0.46 \\
   INTEGRAL      &  1.0$^{+0.4}_{-0.4}$ & 1.11$\pm$0.15 & 0.98--1.36\\
   Suzaku (1st)  & 1.30$^{+0.07}_{-0.14}$ & 2.7$^{+0.2}_{-0.3}$ & 1.39 \\
   Suzaku (2nd)  & 1.51$^{+0.04}_{-0.15}$ & 1.9$^{+0.3}_{-0.1}$ & 1.47 \\
   Suzaku (both) & 1.38$^{+0.14}_{-0.13}$ & 2.3$^{+0.7}_{-0.6}$ & 1.39, 1.47\\
   \hline
   \multicolumn{4}{@{}l@{}}{\hbox to 0pt{\parbox{85mm}{\footnotesize
   \par\noindent
   \footnotemark[$*$] The best-fit photon index ($\Gamma$) and the flux ($F_{\rm{X}}$)
   in the 22--100~keV range of the power-law model.
   \par\noindent
   \footnotemark[$\dagger$] The ranges indicate the 90\% statistical
   uncertainty. The uncertainty for the INTEGRAL value was derived from \citet{leyder08}.
   \par\noindent
   \footnotemark[$\ddagger$]
   The orbital phase 1.0 indicates the onset of the Cycle 11 minimum at the Julian date
   2452819.8 d \citep{damineli08}.
   }\hss}}
  \end{tabular}
 \end{center}
\end{table}  

We compare the results for the super-hard excess emission among the two Suzaku
observations and the two previous studies. We use the spectral parameters for the
power-law model, which are available both for the BeppoSAX and the INTEGRAL results
(table~\ref{t5}). The flux values of BeppoSAX \citep{viotti04} and Suzaku
(table~\ref{t3}) were reevaluated in the 22--100~keV range to facilitate comparison with
the INTEGRAL result \citep{leyder08}.

We see no significant change between the two Suzaku results both for the flux and the
power-law index if we consider both the statistical and systematic
uncertainties. Therefore, we fitted the two 5--50~keV spectra simultaneously to obtain a
much constrained fit. In the fitting, the best-fit parameters of the thin-thermal plasma
components were fixed to the best-fit values for the independent fits. The parameters of
the statistically-acceptable best fit are appended in table~\ref{t5}.

The Suzaku results were obtained at different epochs during the X-ray steady increase
phase near the apastron with no apparent flare-like flux amplification in the hard
band. This infers that the super-hard excess emission is not caused by episodic events,
but is a constant emission with little change both in the flux and the spectral
hardness.

The best-fit Suzaku model is compared to the INTEGRAL spectrum in
figure~\ref{f6}. $\Gamma$ is the same within the statistical and the systematic
uncertainties, but $F_{\rm{X}}$ is significantly higher in Suzaku than INTEGRAL
(table~\ref{t5}). The INTEGRAL result was obtained in a wide range of orbital phases
including the minimum, recovery, and steady increase by accumulating a 1~Ms exposure. We
speculate that the lower flux in the INTEGRAL data is a consequence of the fact that the
INTEGRAL spectrum was accumulated in periods including the orbital eclipse, during which
the super-hard emission may be absent.

On the other hand, the Suzaku and BeppoSAX \citep{viotti04} results are inconsistent
$\Gamma$, which may be due to the contamination to the PDS spectra by the neighboring
sources \citep{leyder08}.

\subsection{Origin of the Super-hard Excess Emission}
The emission mechanism of the super-hard excess X-rays and the production mechanism of
the charged particles (mostly electrons) to produce the radiation are two different
problems. For the emission mechanism, we consider the bremsstrahlung and the inverse
Compton radiations. The bremsstrahlung emission is either thermal stemming from
thermally-relaxed high-temperature plasma or non-thermal from accelerated charged
particles with a non-thermal energy distribution.

First, we consider the thermal bremsstrahlung by a thermally-relaxed plasma. In an
adiabatic strong shock, the post shock temperature is $3 \mu m_{\rm{H}}
v^{2}_{\rm{shock}}/16 k_{\rm{B}}$, where $\mu$ is the average particle mass in the unit
of the proton mass ($m_{\rm{H}}$) and $v_{\rm{shock}}$ is the shock
velocity. Substituting $v_{\rm{shock}}=$3000~km~s$^{-1}$ \citep{pittard02}, we obtain
$T=$~18$\mu$~keV. $\mu$ increases for H-deficient gas widely seen among stellar winds
from evolved early-type stars such as $\eta$ Car, and reaches 4/3 for the most extreme
case that the gas is composed only of fully-ionized He. The thermal bremsstrahlung model
was abandoned in the BeppoSAX \citep{viotti04} analysis and was unemployed in the
INTEGRAL analysis \citep{leyder08}. The combined Suzaku spectrum indicates
$k_{\rm{B}}T^{\rm{(br)}} >$~35~keV, but given a large systematic uncertainty, we
consider that this emission mechanism remains a viable explanation.

Second, we argue against the bremsstrahlung emission by accelerated charged particles
colliding upon ambient cold matter. This process is proposed for the super-hard emission
seen in solar and stellar flares \citep{osten07}. $\eta$ Car has a rich ambient matter
to brake the accelerated particles, which is evidenced by a large amount of
circumstellar extinction and the presence of the Fe\emissiontype{I} fluorescence line
\citep{corcoran04}. However, the X-ray yield of this process is very low of
$\approx$10$^{-5}$ even for the most efficient bremsstrahlung by electrons of a 10~keV
to 1~MeV kinetic energy against a thick target. Most of the kinetic energy is lost by
the ionization of cold target. The measured luminosity is $\sim 1 \times
10^{34}$~erg~s$^{-1}$ in the super-hard band for the two Suzaku observations, which
implies that a constant energy input is necessary at a rate of
$\approx$10$^{39}$~erg~s$^{-1}$. This is hardly possible given the total kinetic energy
of the stellar wind in this CWB being $\sim$10$^{37}$~erg~s$^{-1}$ \citep{leyder08}.

Finally, we consider the inverse Compton emission of stellar UV photons by a non-thermal
population of electrons scattering the ambient UV photons up to the super-hard band
\citep{chen91}. This is the most popular explanation adopted in \citet{viotti04} and
\citet{leyder08}, and our Suzaku result agrees with this interpretation. Several
mechanisms of acceleration to produce the non-thermal population are proposed, which
includes the first-order Fermi acceleration at the shock, electric field acceleration in
the current sheet in a magnetized wind \citep{jardine96}, and magnetic
reconnections. The X-ray photon index of $\sim$1.4 indicates that the injected
non-thermal electrons have an energy distribution with an index of $\sim$2 or harder,
depending on how the Compton cooling balances the injected energy. The diffusive shock
acceleration predicts the value of 2 for the electron energy index, and some
modifications may need to be considered \citep{pittard06}.

\bigskip

The authors acknowledge Y. Hyodo for his help in the XIS data reduction, J. Pittard for
his comments on particle acceleration, and J. C. Leyder and R. Walter for providing the
INTEGRAL spectrum. We thank the Suzaku science working group team for their effort in
the initial phase of the operation, during which the present data set was
obtained. Support for this work is provided by the Research Center of the Advanced
Measurement at Rikkyo University (A.\,S.), the Grants-in-Aid for Scientific Research by
the Ministry of Education, Culture, Sports, Science and Technology of Japan (grant
numbers 19654032 and 19340047 for S.\,K. and 20540237 for Y.\,T.), a Chuo University
Grant for Special Research (Y.\,T.), and the National Aeronautics and Space
Administration through Chandra Postdoctoral Fellowship Award Number PF6-70044 (M.\,T.)
issued by the Chandra X-ray Observatory Center, which is operated by the Smithsonian
Astrophysical Observatory for and on behalf of the National Aeronautics Space
Administration under contract NAS8-03060.

\end{document}